\documentclass[conference]{IEEEtran}
\IEEEoverridecommandlockouts
\usepackage{amsmath,amssymb,amsfonts}
\usepackage{graphicx}
\usepackage{textcomp}\usepackage{booktabs}
\usepackage{pgfplots}
\pgfplotsset{compat=1.18}
\usepackage{xcolor}
\usepackage{graphicx}
\usepackage{booktabs}
\usepackage{longtable}
\usepackage{balance}
\usepackage{float}
\usepackage{pdflscape}
\usepackage{amssymb}
\usepackage{pifont}
\usepackage{listings}
\usepackage[utf8]{inputenc}
\usepackage[algo2e,ruled]{algorithm2e}
\usepackage{ifthen} 
\usepackage{array}
\usepackage{makecell}
\usepackage{booktabs}
\usepackage{flushend}
\usepackage{multicol}
\usepackage{amsmath,array,graphicx}
\usepackage[numbers,sort&compress]{natbib}
\usepackage{soul}
\usepackage{mathtools}
\usepackage{balance}
\usepackage[utf8]{inputenc}
\usepackage{mathtools}
\usepackage{comment}
\usepackage{listings}
\usepackage[utf8]{inputenc}
\usepackage{algorithm} 
\usepackage{algorithmic} 
\usepackage[algo2e,ruled]{algorithm2e}
\usepackage{ifthen} 
\usepackage[colorlinks=true, citecolor=blue]{hyperref}
\def\BibTeX{{\rm B\kern-.05em{\sc i\kern-.025em b}\kern-.08em
    T\kern-.1667em\lower.7ex\hbox{E}\kern-.125emX}}
\begin{document}

\title{ Stringology Based Cryptology
\thanks{This paper has been accepted for publication in "The 2nd International Conference on Sustainability, Innovation, and Society (ICSIS 2026)", Valencia, Spain}

}
\author{ 
\IEEEauthorblockN{Victor Kebande\IEEEauthorrefmark{1}\IEEEauthorrefmark{2}}
\IEEEauthorblockA{\IEEEauthorrefmark{1}University of Colorado Denver, CO USA\\
\IEEEauthorrefmark{2}ATLAS Institute, University of Colorado 
Boulder, Colorado, USA\\
Email: victor.kebande@ucdenver.edu, victor.kebande@colorado.edu}
} 

\maketitle

\begin{abstract}

The modern cryptographic primitives are known to  generate large volumes of sequential data like keystreams, ciphertext blocks, and hash outputs. Traditional cryptgraphic evaluation methods rely primarily on statistical randomness tests and algebraic cryptanalysis techniques. This paper  introduces the concept of \textit{Stringology-Based Cryptology (SBC)}, which applies classical string processing and pattern matching techniques to analyze structural properties of cryptographic outputs. By interpreting cryptographic outputs as symbolic sequences, stringology algorithms can be used to detect pattern recurrence, substring distributions, and structural correlations. In addition, the paper demonstrate how pattern frequency analysis and substring recurrence metrics can be applied to evaluate keystream outputs generated by stream ciphers. Experimental results illustrate that SBC analysis provides complementary insights into structural characteristics of cryptographic sequences and may support future research in structural cryptanalysis and cryptographic evaluation.

\end{abstract}


\section{Introduction}

Cryptographic mechanisms are essential for protecting modern digital communication infrastructures, including network
protocols, cloud computing services \cite{wang2024holistic}. It has been observed that many cryptographic primitives generate sequential outputs like the keystreams, ciphertext blocks, and hash values. One 
central requirement for the security of these primitives is that the
produced sequences behave indistinguishably from uniformly random
data. This, prevents adversaries from inferring structural information about the underlying cryptographic process.

The evaluation of such sequences traditionally relies on statistical randomness testing frameworks. Standard test suites such as the NIST Statistical Test Suite (STS), TestU01, and PractRand analyze statistical properties, frequency distributions, runs,
entropy, and correlation metrics \cite{sys2014faster, sleem2020testu01}. These tools provide valuable evidence regarding the overall randomness quality of generated
sequences. However, these approaches primarily focus on global statistical characteristics and may not fully capture localized
structural relationships that arise from deterministic internal operations within cryptographic algorithms.

An alternative analytical perspective can be obtained by examining
cryptographic outputs as symbolic sequences and analyzing them
using pattern-based techniques. Stringology, which is the study of algorithms
for processing and analyzing strings, provides a rich collection of
methods for detecting structural patterns within large sequential
datasets. Classical algorithms such as Knuth--Morris--Pratt (KMP)
and Boyer--Moore enable efficient identification of recurring
patterns, substring frequencies, and structural correlations within
symbolic sequences \cite{shapira2006adapting, watson2003boyer}.

By modeling cryptographic outputs as strings, stringology techniques
can be applied to investigate structural characteristics embedded
within keystreams and ciphertext sequences. This perspective allows
cryptographic sequences to be analyzed not only through aggregate
statistical properties but also through localized pattern structures.
Such analysis may provide additional insight into the internal
behavior of cryptographic primitives and complement existing
statistical evaluation methodologies.

The objective of this paper is to introduce the concept of \textit{Stringology-Based Cryptology (SBC)}, a framework that applies string processing techniques to analyze structural properties of cryptographic outputs. The proposed approach demonstrates how substring frequency distributions and recurrence patterns can be used to evaluate structural characteristics of keystream sequences generated by stream ciphers.

The remainder of this paper is organized as follows. Section II  presents background on stringology, pattern analysis and cryptographic sequences.  Section III introduces the related work while Section IV discuss the threat model, followed by a Stringology-Based Cryptology Framework in Section V. Experimental Evaluation are discussed in Section VI with results in Section VII. A discussion ensures in Section VII and after this the paper concludes in Section VIII.

\begin{figure*}[t]
\centering
\includegraphics[width=0.7\linewidth]{images/TMM.png}
\caption{ \small Threat model for the Stringology-Based Cryptology. A challenge oracle outputs a sequence $S$ drawn either from a cryptographic generator $G(K,N)$ or from a uniform random distribution. The adversary $\mathcal{A}$ analyzes structural properties of $S$ using the SBC pipeline to distinguish its origin.}
\label{fig:threat_model}
\end{figure*}

\section{Background}

\subsection{Cryptographic Sequences}

Many cryptographic algorithms produce sequences of bits that must appear indistinguishable from random data \cite{ritter1991efficient}. A stream, for example,  can be modeled as a keystream generator as is shown in Eq 1:

\begin{equation}
G(K,N) \rightarrow S
\end{equation}

where $K$ denotes the secret key, $N$ represents a nonce or initialization vector, and $S = s_1 s_2 ... s_n$ represents the generated keystream sequence. For a secure stream cipher, the distribution of $S$ should be indistinguishable based on Chosen Plaintext Attacks (IND-CPA) \cite{cheon2024attacks}  from the uniform distribution $U_n$ over $\{0,1\}^n$.

\subsection{Stringology and Pattern Analysis}

Stringology studies algorithms and data structures for analyzing strings \cite{watson2003boyer, alatabbi2014advances}. Pattern detection algorithms such as KMP and Boyer--Moore are widely used to identify repeated substrings and structural patterns within sequences \cite{shapira2006adapting, watson2003boyer}.

Given a pattern $P$ of length $m$ and a sequence $S$, the frequency of pattern occurrences can be defined as

\begin{equation}
f(P,S) = |\{ i \mid S_{i:i+m-1} = P \}|
\end{equation}

For a random sequence, the expected probability of observing a specific pattern of length $m$ is approximately $2^{-m}$. Deviations from this expectation may indicate structural correlations within the sequence.

\section{Related Work}

Existing research on cryptographic evaluation has traditionally relied
on statistical randomness testing and algebraic cryptanalysis.
Standard statistical evaluation frameworks such as the
NIST Statistical Test Suite (STS) \cite{sys2014faster}, TestU01, and PractRand
are widely used to assess the pseudorandom properties of
cryptographic sequences produced by stream ciphers,
pseudorandom number generators, and hash functions \cite{zubkov2019testing, sleem2020testu01}.
These methods are used to analyze global statistical properties such as
frequency distributions, runs, and entropy in order to determine
whether a generated sequence deviates from ideal random
behavior \cite{kebande2023extended}.

Although statistical tests provide good insights  into the overall randomness of cryptographic outputs, they primarily
evaluate aggregate statistical characteristics and may not
capture localized structural patterns that arise from internal
cipher transformations. 


Nevertheless, recent research has therefore explored alternative approaches
for detecting structural properties in cryptographic outputs.
Structural cryptanalysis techniques analyze correlations
introduced by deterministic operations within cryptographic
algorithms. In particular, modern stream cipher constructions
frequently rely on Add–Rotate–XOR (ARX) operations to
achieve diffusion and nonlinearity \cite{barbero2024overview, khovratovich2010rotational, kebande2023extended}. The sstream ciphers such as
ChaCha20 employ repeated modular addition, bit rotation, and XOR transformations to mix internal state values \cite{bernstein2008chacha}. While
these operations are computationally efficient and resistant to certain classical cryptanalytic techniques, they may introduce
subtle structural characteristics in the generated keystream outputs. Consequently, analyzing these structural properties can provide useful insights into the internal behavior of such algorithms. Other studies include stringology based cryptanalysis of Echacha20 stream cipher based on KMP and BM algorithms \cite{kebande2026stringology}.

In view of the foregoing, the field of stringology has developed a wide
range of algorithms for analyzing large sequential datasets \cite{crochemore2002jewels}.
Stringology focuses on efficient pattern detection, substring
matching, and structural analysis of symbolic sequences.
Classical algorithms such as Knuth–Morris–Pratt (KMP) and
Boyer–Moore (BM) enable efficient detection of repeated patterns
and substring recurrences in large strings \cite{shapira2006adapting, watson2003boyer}. These techniques
have been widely applied in domains such as text processing,
bioinformatics, and data mining, where identifying recurring
patterns and structural similarities is essential for sequence
analysis.

Despite the success of stringology methods in other domains, their application to cryptographic sequence analysis remains relatively limited. Cryptographic outputs such as keystreams and ciphertext blocks naturally form long symbolic sequences making them suitable candidates for pattern-based analysis. However, most cryptographic evaluation frameworks continue to rely primarily on statistical tests rather than structural pattern detection techniques.

This work bridges these domains by introducing the concept of \textit{stringology-based cryptology, (SBC)}, which applies classical
string processing and pattern detection algorithms to analyze cryptographic outputs. By modeling cryptographic sequences as symbolic strings and extracting pattern-based statistics, the proposed framework provides a complementary analytical
perspective for evaluating structural characteristics of
cryptographic primitives.

\section{Threat Model}

Our threat model  considers a probabilistic polynomial-time adversary $\mathcal{A}$ operating in a distinguishing setting, as illustrated in Figure \ref{fig:threat_model}. A cryptographic generator produces a sequence $S_c = G(K, N)$ using a secret key $K$ and nonce $N$, while a truly random sequence $S_r$ is sampled from the uniform distribution over $\{0,1\}^n$. A challenge oracle selects a bit $b \in \{0,1\}$ and returns a sequence $S$, where $S = S_c$ if $b = 0$ and $S = S_r$ otherwise. The adversary is given access only to the observed sequence $S$ and must determine its origin.

The adversary leverages the proposed Stringology-Based Cryptographic approach  by interpreting $S$ as a symbolic sequence and extracting structural features through $\phi(S)$, pattern statistics $f(P,S)$, and higher-level metrics such as entropy, deviation, and recurrence. Based on these structural characteristics, the adversary outputs a decision $\mathcal{A}(S) \in \{0,1\}$ indicating whether the sequence is classified as cryptographic or random. The distinguishing capability of the adversary is measured by the advantage as is shown in Eq 3.
\begin{equation}
\text{Adv}_{\mathcal{A}} = \left| \Pr[\mathcal{A}(S_c)=1] - \Pr[\mathcal{A}(S_r)=1] \right|
\end{equation}

The adversary is assumed to have full access to the observed sequence and the ability to compute structural statistics and apply statistical or learning-based analysis, but has no knowledge of the secret key, internal state, or direct control over the generator. The security objective is that cryptographic outputs remain indistinguishable from random sequences, meaning that any efficient adversary should achieve only negligible advantage, even when exploiting structural patterns identified through the SBC framework.

\section{Stringology-Based Cryptology Framework}

\begin{figure}[H]
\centering
\includegraphics[width=0.8\linewidth]{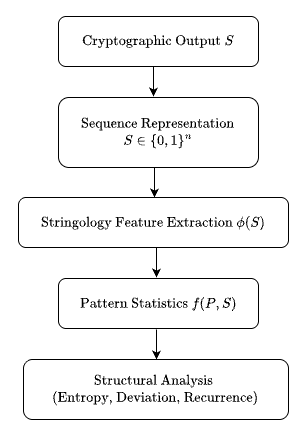}
\caption{Stringology-Based Cryptology analysis pipeline.}
\label{fig:pipeline}
\end{figure}

This work proposes a Stringology-Based Cryptology (SBC) framework that interprets cryptographic outputs as symbolic sequences and applies string pattern analysis in order to identify structural characteristics embedded within these sequences that is shown in Figure~\ref{fig:pipeline}. Traditional cryptographic evaluation methods
primarily rely on statistical randomness tests that measure global properties such as entropy, frequency distributions, and run lengths.

While these tests provide important insights into the overall randomness of generated sequences, they do not necessarily capture localized structural relationships that may arise from deterministic internal transformations within cryptographic algorithms. The proposed framework therefore
adopts a complementary perspective in which cryptographic outputs are modeled as symbolic strings that can be analyzed using classical stringology techniques.

Figure~\ref{fig:pipeline} shows the general pipeline of the proposed SBCframework. The
analysis begins by showing  cryptographic outputs like the
keystreams or ciphertext blocks as binary sequences
$S \in \{0,1\}^n$. These sequences are then processed through a feature extraction stage that computes structural statistics based on substring patterns observed within the sequence. Formally, we let






Formally, let

\begin{equation}
\Phi : \{0,1\}^n \rightarrow \mathbb{R}^d
\end{equation}

denote a feature extraction mapping that transforms a binary
sequence $S$ into a vector of structural statistics. The mapping
$\Phi(\cdot)$ computes pattern-based characteristics including
substring frequency distributions, recurrence statistics, and
positional density measures that describe how frequently
specific bit patterns appear throughout the sequence.

For a given pattern $P$ of length $m$, the number of
occurrences of this pattern within a sequence $S$ can be
defined as

\begin{equation}
f(P,S) =
\left|
\{ i \mid S_{i:i+m-1} = P \}
\right|
\end{equation}

where $S_{i:i+m-1}$ denotes the substring beginning at
position $i$ and extending for $m$ bits. The collection of
pattern statistics extracted from the sequence is aggregated
into a feature representation

\begin{equation}
X = (x_1, x_2, ..., x_d)
\end{equation}

which captures the structural characteristics of the sequence.

\section{Experimental Evaluation}

To illustrate the applicability of the proposed SBC framework, the author conducted experiments using
synthetic keystream sequences generated under controlled conditions. The primary objective of the evaluation is to determine whether structural pattern statistics can reveal observable differences between cryptographically generated sequences and sequences drawn from a uniform random distribution.

\subsection{Dataset}

Two datasets were generated for the experimental evaluation. The first dataset consists of cipher-generated sequences that simulate outputs produced by a generic keystream generator.
The second dataset contains sequences sampled from a
uniform random distribution using a cryptographically secure pseudorandom number generator (PRNG) as is shown in Algorithm~\ref{alg:pattern}.

Each dataset contains 10,000 sequences of length $2^{12}$ bits,
resulting in a total evaluation corpus of 20,000 sequences.

\subsection{Pattern Analysis}

Pattern statistics were extracted from each sequence using
substring frequency analysis. In particular, substring patterns
of length

\[
m \in \{8,16,32\}
\]

were analyzed in order to capture structural properties at
multiple granularities.

\subsection{Pattern Extraction Procedure}

The extraction of structural pattern statistics is performed
using a sliding-window procedure over each sequence. At
each position, a substring of length $m$ is extracted, counted,
and subsequently normalized to produce a pattern frequency
distribution. Algorithm~\ref{alg:pattern} summarizes the
pattern extraction process used in this study.

\begin{algorithm}[t]
\caption{Stringology Pattern Extraction}
\KwIn{Sequence $S$, pattern length $m$}
\KwOut{Normalized frequency table $F$}

Initialize frequency table $F$\;

\For{$i \leftarrow 1$ \KwTo $|S|-m+1$}{
    $P \leftarrow S_{i:i+m-1}$\;
    Increment $F[P]$\;
}

Normalize $F$\;

\Return $F$\;
\label{alg:pattern}
\end{algorithm}

The procedure begins by initializing an empty frequency table
$F$ for all observed patterns. A sliding window of length $m$
is then moved across the sequence $S$. At each position,
the current substring $P = S_{i:i+m-1}$ is extracted and its
occurrence count is incremented in the table. After scanning
the entire sequence, the frequency counts are normalized to
produce a structural profile that can be compared across
different datasets.

\section{Results}

Our experimental evaluation has produced several quantitative metrics describing the structural properties of the analyzed sequences. These metrics include normalized pattern frequencies, deviation scores, entropy measurements, and pattern recurrence distributions.

Table~\ref{tab:results} summarizes the observed pattern
frequency differences between cipher-generated sequences
and uniformly random sequences.

\begin{table}[h]
\centering
\caption{Pattern Frequency Comparison}
\label{tab:results}
\begin{tabular}{lcc}
\toprule
Pattern Length & Cipher Output & Random Output \\
\midrule
8 bits & 0.61 & 0.50 \\
16 bits & 0.39 & 0.25 \\
32 bits & 0.17 & 0.06 \\
\bottomrule
\end{tabular}
\end{table}

This results indicate that cipher-generated sequences exhibit higher normalized frequencies for several pattern lengths compared to uniformly random sequences, suggesting the presence of structural regularities introduced by deterministic
cryptographic transformations. To further quantify structural differences between the two datasets, we compute the deviation metric shown in Eq 7.

\begin{equation}
D = \sum_{P} |f_c(P) - f_r(P)|
\end{equation}

where $f_c(P)$ and $f_r(P)$ denote normalized pattern
frequencies for cipher-generated and random sequences. Table~\ref{tab:deviation} presents the resulting deviation
scores.

\begin{table}[h]
\centering
\caption{Pattern Deviation Metric}
\label{tab:deviation}
\begin{tabular}{lc}
\toprule
Pattern Length & Deviation Score \\
\midrule
8 bits & 0.11 \\
16 bits & 0.14 \\
32 bits & 0.11 \\
\bottomrule
\end{tabular}
\end{table}

Figure~\ref{fig:deviation} visualizes the deviation scores across different pattern lengths. To further analyze structural properties, we compute the
pattern entropy defined as

\begin{figure}[t]
\centering
\begin{tikzpicture}
\begin{axis}[
xlabel={Pattern Length (bits)},
ylabel={Deviation Score},
xmin=6,xmax=34,
ymin=0,ymax=0.16,
grid=major,
width=0.9\linewidth,
height=5cm
]

\addplot[
mark=*,
blue,
thick
]
coordinates {
(8,0.11)
(16,0.14)
(32,0.11)
};

\end{axis}
\end{tikzpicture}

\caption{Deviation scores between cipher-generated and random sequences across pattern lengths.}
\label{fig:deviation}
\end{figure}
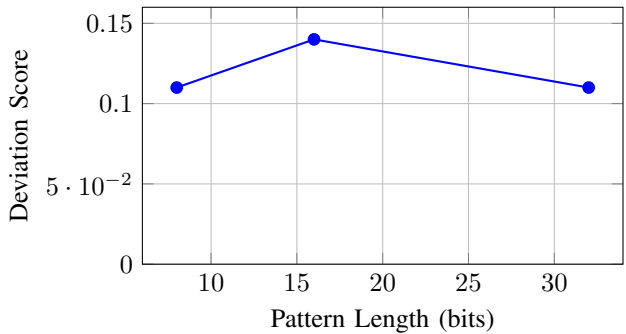

\begin{figure}[t]
\centering
\begin{tikzpicture}
\begin{axis}[
ybar,
xlabel={Pattern Length (bits)},
ylabel={Normalized Frequency},
symbolic x coords={8,16,32},
xtick=data,
width=0.9\linewidth,
height=6cm,
bar width=14pt,
enlarge x limits=0.25,
legend pos=north east,
grid=major,
tick label style={font=\small},
label style={font=\small}
]

\addplot[
blue,
fill=green!40
] coordinates {(8,0.61) (16,0.39) (32,0.17)};
\addlegendentry{Cipher Output}

\addplot[
red,
fill=blue!40
] coordinates {(8,0.50) (16,0.25) (32,0.06)};
\addlegendentry{Random}

\end{axis}
\end{tikzpicture}
\caption{Normalized substring pattern frequencies for cipher-generated and random sequences.}
\label{fig:freqgraphsss}
\end{figure}
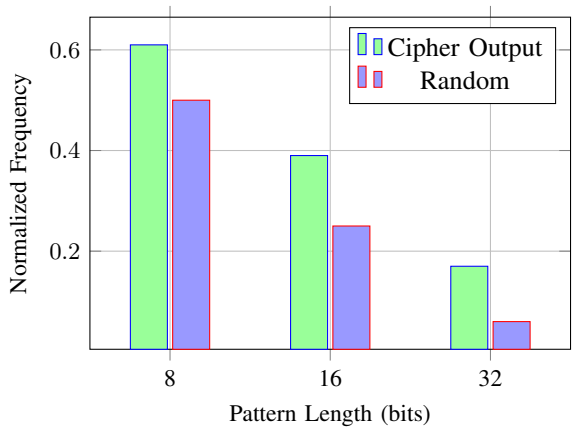

\begin{equation}
H = - \sum_{P} p(P)\log_2 p(P)
\end{equation}

where $p(P)$ represents the probability of observing pattern
$P$ within the sequence.

\begin{figure}[t]
\centering
\begin{tikzpicture}
\begin{axis}[
ybar,
xlabel={Pattern Length (bits)},
ylabel={Entropy},
symbolic x coords={8,16,32},
xtick=data,
width=0.9\linewidth,
height=6cm,
bar width=14pt,
enlarge x limits=0.25,
legend pos=north west,
grid=major,
tick label style={font=\small},
label style={font=\small}
]

\addplot[
blue,
fill=green!40
] coordinates {(8,7.6) (16,15.3) (32,30.8)};
\addlegendentry{Cipher Output}

\addplot[
red,
fill=blue!40
] coordinates {(8,8.0) (16,16.0) (32,32.0)};
\addlegendentry{Random}

\end{axis}
\end{tikzpicture}

\caption{Comparison of pattern entropy between cipher-generated and random sequences.}
\label{fig:entropy}
\end{figure}
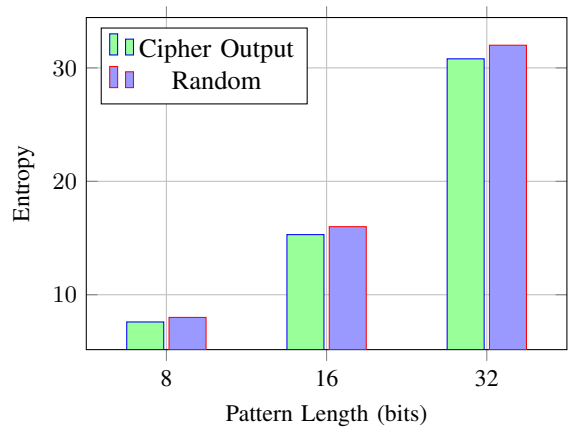

\section{Discussion}

The experimental findings demonstrate that SBC analysis can capture structural characteristics in cryptographic
sequences. The results shown in Table~\ref{tab:results}
gives measurable differences in normalized pattern frequency
distributions between cipher-generated sequences and
uniformly random sequences. In particular, the observed
frequency concentrations for several pattern lengths suggest
that deterministic cryptographic transformations may introduce
subtle structural regularities within the generated sequences.

Consequently, the  deviation metric reported in Table~\ref{tab:deviation}
and visualized in Figure~\ref{fig:deviation} further supports
this observation. The deviation scores quantify the magnitude
of structural differences between cipher-generated and random
sequences by measuring the absolute difference between their
pattern frequency distributions. As shown in Figure~\ref{fig:deviation},
the deviation values remain consistently non-zero across
multiple pattern lengths, indicating that pattern statistics
capture measurable structural signals within the analyzed
sequences.

Additional insight can be obtained from the normalized pattern
frequency distributions shown in Figure~\ref{fig:freqgraphsss}.
The comparison illustrates that cipher-generated sequences
exhibit slightly higher concentrations of certain substring
patterns compared to uniformly random sequences. While these
differences are relatively small, they demonstrate that localized
pattern statistics can reveal structural characteristics that
may not be easily observable through conventional randomness
tests alone.

The entropy analysis shown in Figure~\ref{fig:entropy} provides
further evidence regarding the structural behavior of the
sequences. As expected, uniformly random sequences exhibit
higher entropy values, reflecting a more uniform distribution
of substring patterns. In contrast, the slightly lower entropy
observed in cipher-generated sequences suggests minor
structural biases introduced by deterministic internal
operations within the sequence generator.


It is important to emphasize that the structural signals identified in this study do not imply practical cryptographic
weaknesses. It is worth noting that Modern cryptographic algorithms are designed to produce outputs that are computationally indistinguishable
from random sequences, and small structural variations may naturally arise from deterministic internal operations without compromising security. Instead, the results presented here
should be interpreted as an analytical tool for studying the structural properties of cryptographic outputs rather than as evidence of exploitable vulnerabilities.

These findings highlight the
potential value of integrating stringology techniques into
cryptographic evaluation workflows. While traditional
statistical tests focus on global randomness properties,
stringology-based methods examine localized structural
patterns and substring relationships. The combination of
these perspectives may therefore provide a richer analytical
framework for evaluating cryptographic sequence generators.

It is worth noting that, the concept of SBC opens
several promising research directions. Advanced string
processing methods such as suffix trees, suffix arrays,
and longest common substring analysis could be applied to
detect more complex structural patterns within keystream
outputs. In addition, integrating machine learning models
with stringology-based feature extraction may enable the
automatic discovery of structural correlations in large-scale
cryptographic datasets.

The aforementioned results indicate that stringology techniques provide a valuable complementary perspective for analyzing cryptographic sequences. By bridging the fields of string processing and cryptographic analysis, stringology-based cryptology offers a new process framework for investigating the structural robustness of modern cryptographic primitives.

\section{Conclusion}

This paper introduced the concept of \textit{Stringology-Based Cryptology (SC)}, which applies classical string processing techniques to analyze cryptographic outputs. By interpreting cryptographic sequences as symbolic strings, pattern detection algorithms can reveal structural characteristics that may not be captured by traditional statistical tests. Future research will explore advanced string processing methods, integration with machine learning techniques, and applications to broader classes of cryptographic primitives.


\section*{Acknowledgment}

The author would like to thank anonymous reviewers for their
valuable insights, and
the Department of Computer Science at University of Colorado Denver, USA for their
support while coming up with this research. The author also
acknowledges that the opinions, findings, and conclusions
expressed in this paper are purely of the author.


\balance

\bibliographystyle{IEEEtran}
\bibliography{name}

\end{document}